\documentclass[10pt]{iopart}
\usepackage[utf8]{inputenc} 
\usepackage{graphicx}
\usepackage{amsfonts}
\newcommand{\be}{\begin{equation}}
\newcommand{\ee}{\end{equation}}
\newcommand{\bea}{\begin{eqnarray}}
\newcommand{\eea}{\end{eqnarray}}

\newcommand{\eqref}[1]{(\ref{#1})}
\usepackage{color}

\usepackage[colorinlistoftodos]{todonotes}

\usepackage[margin=1.25in]{geometry}

\begin{document}

\title{Anisotropic collisions of dipolar Bose-Einstein condensates in the universal regime}

\author{Nathaniel Q. Burdick$^{1,2}$, Andrew G. Sykes$^{3}$, Yijun Tang$^{2,4}$, and Benjamin L. Lev$^{1,2,4}$}
\address{$^1$ Department of Applied Physics, Stanford University, Stanford CA 94305, USA}
\address{$^2$ E.~L.~Ginzton Laboratory, Stanford University, Stanford CA 94305, USA}
\address{$^3$ LPTMS, CNRS, Univ.~Paris Sud, Universit\'e Paris-Saclay, 91405 Orsay, France}
\address{$^4$ Department of Physics, Stanford University, Stanford CA 94305, USA}
\ead{benlev@stanford.edu}

\date{\today}
\begin{abstract}
We report the measurement of collisions between two Bose-Einstein condensates with strong dipolar interactions. The collision velocity is significantly larger than the internal velocity distribution widths of the individual condensates, and thus, with the condensates being sufficiently dilute, a halo corresponding to the two-body differential scattering cross section is observed. The results demonstrate a novel regime of quantum scattering, relevant to dipolar interactions, in which a large number of angular momentum states become coupled during the collision. We perform Monte-Carlo simulations to provide a detailed comparison between theoretical two-body cross sections and the experimental observations.
\end{abstract}


\section{Introduction}

In general, the description of quantum mechanical scattering becomes simpler when the relative kinetic energy between the collision partners is negligibly small.  Specifically, in this limit the de Broglie wavelength of relative motion greatly exceeds the length scale over which the two particles exert forces on one another, rendering details of their interaction, if not insignificant, at least fairly simple to account for~\cite{LandauQuantum}.  This concept was first articulated by Enrico Fermi in the context of collisional broadening of spectral lines in a gas of Rydberg atoms \cite{Fermi34_NC}; was instrumental in the understanding of low-energy scattering of neutrons from nuclei \cite{Bethe}; and has found expression in modern times as the bedrock upon which our understanding of ultracold gases is based \cite{Chin10_RMP}. Ordinarily, the key aspect of this simplicity is that the scattered state contains only a single (or very few) eigenstate(s) of angular momentum, i.e., one or only a few partial waves. Ultracold identical bosons have a differential cross 
section which is isotropic and independent of the collision energy such that a single length scale, referred to as the scattering length, suffices for its description. The situation is only marginally more complicated for ultracold identical fermions in which the differential cross section is $\propto k^4|\hat{k}\cdot\hat{k}'|^2$, where $k$ is the magnitude of the relative wave vector and $\hat{k}$ ($\hat{k}'$) is a unit vector along the direction of relative incoming (outgoing) momentum~\cite{TaylorBook}. Once again, a single scalar quantity, the scattering volume, sufficiently characterises the underlying potential. These simple results dictate many important dynamical properties of quantum gases such as the efficiency of evaporative cooling~\cite{PhysRevLett.80.3419,PhysRevA.59.1500,DeMarcoJin1999_PRL,DeMarcoJin_Science1999} and the speed of sound~\cite{Pethick2002}. 

However, this situation is radically altered when the potential energy between the collision partners does not decay suitably fast. If the interaction potential decays as $1/r^n$, where $r$ is the distance separating the two particles, then for $n\leq3$ a different scenario emerges~\cite{Shakeshaft_r3potential,DipolarCollisions_OMalley}. Ultracold gases containing atoms or molecules which possess  magnetic~\cite{Chromium_BEC,Dysprosium_BEC,Erbium_BEC,Dysprosium_DegenerateFermiGas,Erbium_FermiGas,Chromium_FermiGas} or electric~\cite{KRb,RbCs,NaK,NaRb} dipole moments therefore provide a magnificent opportunity to observe a novel regime of low-energy quantum scattering.  When the dipoles are aligned along a chosen axis by an externally applied field, the overall magnitude of the cross section is still determined by a characteristic length scale, the dipole length, which depends on the dipole moment~\cite{Bohn09_NJP}. However, the dipole alignment direction and the $1/r^3$ asymptotic decay of the interaction potential conspire to create a differential cross section which now involves a very large set of partial waves in an essential way due to the coupling between different angular momentum states.  Moreover, it 
presents a novelty in that such a differential cross section depends explicitly on the relative momenta {\it before} the collision (as well as after) with respect to the polarization axis of the dipoles~\cite{Bohn14_PRA}. The consequence of this dependence has been measured indirectly through its effect on the equilibration rate of a dipolar gas, which is taken out of equilibrium by diabatically changing the trap along a certain direction~\cite{Aikawa14_preprint,TangSykes2015_PRA}. It was seen that the equilibration rate can vary by as much as a factor of four, depending on the angle between the dipole alignment direction and the dynamic axis of the trap.

\begin{figure}[t!]
\centering
 \includegraphics[width=\columnwidth]{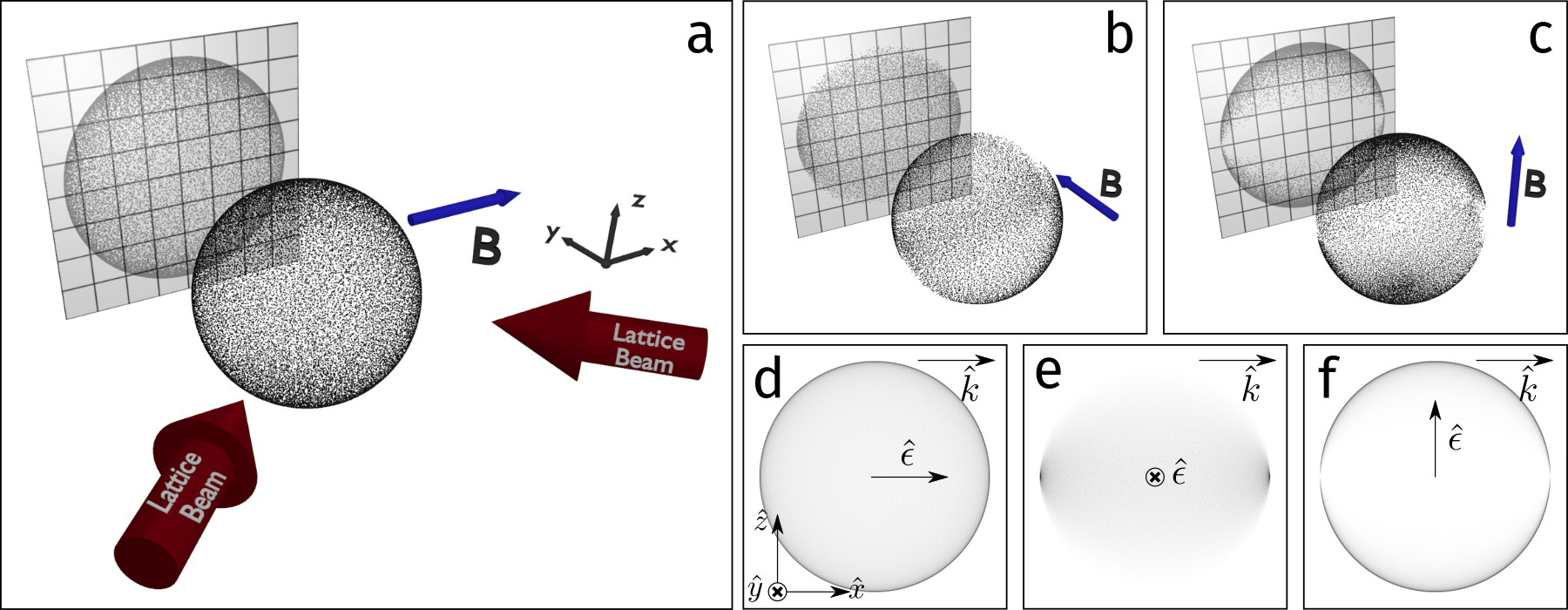}
 \caption{(a) Schematic of BEC lattice diffraction used to produce scattered halos. Two 741-nm beams propagating along $\hat{x}+\hat{y}$ and $\hat{y}-\hat{x}$ produce a lattice along $\hat{x}$ which undergoes a precisely calculated two-pulse sequence to optimally split the BEC in half, with each piece travelling along $\pm\hat{x}$. The 3D scattered halo is measured via absorption imaging along a $\hat{y}$-projection. The dipole alignment is set by a bias magnetic field $B$. (b) and (c) 3D scattered halos and projections with the bias field along $\hat{y}$ and $\hat{z}$, respectively. (d) 2D projection along $\hat{y}$ of the differential cross section calculated from Eq.~(\ref{EQdifferential}) with dipole alignment $\hat{\epsilon}$ along $\hat{x}$. (e) and (f) Same as (d), but with dipole alignment $\hat{\epsilon}$ along $\hat{y}$ and $\hat{z}$, respectively.   The 2D projections presented in (d)-(f) correspond to the halos in panels (a)-(c), respectively.}
    \label{FIG:schematic}
\end{figure}

We report a direct measurement of the differential scattering cross section of $^{162}$Dy, which has an exceptionally large magnetic moment of 9.93$\mu_{\rm B}$, where $\mu_{\rm B}$ is the Bohr magneton.  We obtain this measurement by colliding two Bose-Einstein condensates (BECs) at a relative velocity significantly greater than the width of their respective internal velocity distributions. Furthermore, most of these measurements  are made in the dilute limit where the majority of particles pass through the opposing condensate without experiencing a collision.
In this situation, scattered particles emerge on a spherical halo defined by the constraints of conserved energy and momentum. The angular distribution of the particles on the sphere is almost entirely determined by the differential cross section.
We vary the dipole alignment direction relative to the collision axis and take absorption images of the post-collision number density after a long time-of-flight (TOF). To obtain a detailed theoretical understanding of these images, we perform Monte-Carlo simulations, which capture the essential features of the experiment. 

Related experiments on bosonic alkali gases which also observed halo-like structures due to atomic collisions were reported in Refs.~\cite{swaveHalo_1995,ImpurityScattering_2000,sanddwaveHalo_2004_Niels,sanddwaveHalo_2004_Jook} and recently in fermionic $^{40}$K~\cite{Genkina:2016cb,Niels_Nature}. Remarkably beautiful experiments  were reported in Ref.~\cite{PairCorrelations_MetastableHelium} in which metastable helium was employed. In addition to the halo structure, the internal energy of metastable helium allowed for pair correlations to be measured via spatially and temporally resolved single-atom counting. This stimulated theoretical interest in the strength of such correlations~\cite{PairCorrelations_Theory1,PairCorrelations_Theory2,PairCorrelations_Theory3}. All effects from dipolar interactions are negligible in these experiments involving alkali and He gases, and the reported halo is well described by a very small set of partial waves. A unique experiment reported in Ref.~\cite{Spielman_SyntheticPartialWaves} studied the creation of artificial partial waves in $^{87}$Rb BEC collisions mediated by an optical potential which modifies the constraints due to energy and momentum conservation. Using this approach they were able to engineer differential scattering with several partial waves, in spite of the fact that a bare collision would have been entirely $s$-wave. Our work presents a natural and intriguing extension of these experiments, demonstrating the nature of differential scattering in the presence of dipolar interactions which are anisotropic and unavoidably couple to a large number of angular momentum states during the collision.

\section{Experimental details}

We produce a single BEC  of $^{162}$Dy spin polarized in the $J=8$, $m_J = -8$ absolute ground state with $6.3(4)\times10^{4}$ atoms as described in Ref.~\cite{Tang2015}.  An optical lattice is used to diffract the condensate into the $2n\hbar \mathbf{k}_{L}$ diffraction orders, where the lattice wave vector $\mathbf{k}_{L}=2\pi\sin(\theta/2)/\lambda$ depends on the lattice lasers wavelength $\lambda$ and alignment angle $\theta$, and $n$ is an integer which labels the different diffraction orders. The optical lattice is formed by two fiber-coupled beams derived from the same Ti:Sapphire laser with $\lambda=741$~nm that propagate along $\hat{x}+\hat{y}$ and $\hat{y}-\hat{x}$ such that $\theta=\pi/2$, as shown in Fig.~\ref{FIG:schematic}(a). Both beams are linearly polarized along $\hat{z}$.

The condensate is split into the $\pm2\hbar \mathbf{k}_{L}$ diffraction orders with high efficiency by using a precisely timed two-pulse sequence~\cite{Wu2005}. Thus, two condensates are produced and collide with relative momentum $\mathbf{p}_{\mathrm{rel}}=4\hbar {k}_{L}\hat{x}$.
Immediately after applying the lattice grating, all trapping potentials are removed, and the two condensates collide and expand. After a 22-ms TOF, a column-integrated density profile of the atoms in the $\hat{x}$--$\hat{z}$ plane is measured via absorption imaging along $\hat{y}$. We manipulate the dipole alignment direction by applying a bias magnetic field relative to both the collision axis and the imaging axis. In all configurations, the magnetic field magnitude is held at 1.58~G and away from any Feshbach resonance~\cite{Dy_FeshbachSpectrum}. Three instructive cases are shown in Fig.~\ref{FIG:schematic}, which provides a schematic of our experiment. The differential cross section is isotropic when the dipole alignment is parallel to the relative momentum, as in Fig.~\ref{FIG:schematic}(a). However, with the dipole alignment perpendicular to the relative momentum, as in Fig.~\ref{FIG:schematic}(b) and (c), the differential cross section is clearly anisotropic. While the cross section is the same in panels (b) and (c), the rotation with respect to the imaging direction allows different projections of the 3D halo to be imaged.

\section{Theory and simulation}

\subsection{Two-body scattering theory}~\label{SUBSEC:TwoBody}
The central theoretical element is the differential scattering cross section for identical bosons  interacting via a combination of dipolar and short-range interactions. This is found by solving the two-body Schr\"odinger equation in the center-of-mass rest frame,
\begin{equation}
 \left[-\frac{\hbar^2}{m}\nabla_{\mathbf{r}}^2+V_{\rm d}(\hat{\epsilon},\mathbf{r})+V_{\rm sr}(\mathbf{r})\right]\psi(\mathbf{r})=\frac{\hbar^2k^2}{m}\psi(\mathbf{r}),
\end{equation}
where $m$ is the mass of a single particle and $\hbar$ is the reduced Planck's constant.
The dipole-dipole potential energy for aligned dipoles is $V_{\rm d}(\hat{\epsilon},\mathbf{r})=(2\hbar^2 a_d/m)\left[1-3(\hat{\epsilon}\cdot\hat{r})^2\right]/{r^3}
$,
where $\hat{\epsilon}$ is the direction of alignment.   The potential energy from short-range van der Waals interactions can be approximated by $V_{\rm sr}(\mathbf{r})=(4\pi\hbar^2 a/m)\delta^{(3)}(\mathbf{r})$, provided one stays within the first-order Born approximation.
The strength of the dipole interaction is determined by the length scale $a_d=\mu_0\mu^2m/8\pi\hbar^2$, where $\mu_0$ is the vacuum permeability and $\mu$ is the magnetic moment of a single particle~\cite{DipolarBosonsReview}. Similarly, the strength of the short-range interaction is determined by the $s$-wave scattering length $a$. This problem is solved in detail in Ref.~\cite{Bohn14_PRA}, see also Ref.~\cite{DipolarCollisions_OMalley}. Briefly, in the limit where $k\rightarrow0$, a solution can be found which takes the asymptotic form $\psi(\mathbf{r})=e^{i\mathbf{k}\cdot\mathbf{r}}+f(\hat{\epsilon},\hat{k},\hat{k}')e^{ikr}/r$,
where $e^{i\mathbf{k}\cdot\mathbf{r}}$ is an incoming wave and $f$ is the scattering amplitude that depends on the dipole alignment direction, the incoming wave vector  $\mathbf{k}=k\hat{k}$, and the outgoing wave vector $\mathbf{k}'=k\hat{k}'$.  The differential scattering cross section for identical bosons, given by $\frac{d\sigma}{d\Omega}=\frac{1}{2}|f(\hat{\epsilon},\hat{k},\hat{k}')+f(\hat{\epsilon},\hat{k},-\hat{k}')|^2$ in the low energy limit as $k\rightarrow0$ and within the first-order Born approximation, is found to be
\begin{equation}\label{EQdifferential}
\frac{d\sigma}{d\Omega}=2a_d^2\left[
\frac{(\hat{k}\cdot\hat{\epsilon})^2+(\hat{k}'\cdot\hat{\epsilon})^2-2(\hat{k}\cdot\hat{\epsilon})(\hat{k}'\cdot\hat{\epsilon})(\hat{k}\cdot\hat{k}')}{1-(\hat{k}\cdot\hat{k}')^2}
-\frac{2}{3}+\frac{a}{a_d}
\right]^2,
\end{equation}
where $\hat{k}$ ($\hat{k}'$) is a unit vector along the direction of incoming (outgoing) relative momentum.  The total cross section, found by integrating Eq.~\eqref{EQdifferential} over all possible outgoing directions, is given by~\cite{Bohn14_PRA}
\begin{equation}\label{EQsigma}
 \sigma(\eta) = a_d^2\frac{\pi}{9}\left\{72\frac{a^2}{a_d^2}-24\frac{a}{a_d}\left[1-3\cos^2(\eta)\right]+11-30\cos^2(\eta)+27\cos^4(\eta)\right\},
\end{equation}
where $\eta$ is the angle between $\hat{k}$ and $\hat{\epsilon}$. We note that the spatial anisotropy of this cross section is caused by $\hat{\epsilon}$. Equations~\eqref{EQdifferential} and~\eqref{EQsigma} are universal in the sense that they are insensitive to any details of the short-range potential.

\subsection{Many-body considerations: Monte-Carlo simulation}~\label{SUBSEC:ManyBody}

When two atoms from different condensates collide, their relative velocity is much larger than the width of each condensate's internal velocity distribution. Therefore, these primary collisions effectively occur at a fixed angle between $\hat{k}$ and $\hat{\epsilon}$.  However, each atom then has a finite probability of a secondary scattering event. These secondary scattering events occur with an essentially random angle between $\hat{k}$ and $\hat{\epsilon}$ and corrupt the direct correspondence between the observed halo and the differential scattering cross section. For this reason we use a direct-simulation Monte Carlo (DSMC) algorithm, which is able to keep track of such multiple collision events, in order to quantitatively understand the experimental data as accurately as possible.

The system dynamics can be separated into a low energy part (relevant to the two condensates) and a high energy part (relevant to the halo). These two parts have high and low phase space densities, respectively. Because we are primarily interested in the halo, we focus on the classical kinetic equation for the phase space distribution function, which is given by
\begin{equation}\label{EQ:KineticEq}
    \partial_t f(\mathbf{r},\mathbf{p},t)+\frac{\mathbf{p}}{m}\cdot\nabla_{\mathbf{r}}f(\mathbf{r},\mathbf{p},t)=C[f],
\end{equation}
where the left-hand side contains free-streaming terms and the right-hand side includes effects from two-body collisions. This is given by 
\begin{equation}\label{CollisionIntegral}
C[f]=\int\frac{d^3 \mathbf{p}_1}{h^3}\int d^2\hat{\Omega}\frac{d\sigma}{d\Omega}v_r\left[
f'f_1'-ff_1
\right],
\end{equation}
where $f=f(\mathbf{r},\mathbf{p})$, $f_1=f(\mathbf{r},\mathbf{p}_1)$, $f'=f(\mathbf{r},\mathbf{p}')$, and $f_1'=f(\mathbf{r},\mathbf{p}_1')$ account for the four momenta (two incoming, $\mathbf{p}$, $\mathbf{p}_1$, and two outgoing, $\mathbf{p}'$, $\mathbf{p}_1'$) associated with a two-body collision, and $v_r=|\mathbf{p}-\mathbf{p}_1|/m$ is the relative velocity. Note that $\mathbf{p}=\hbar \mathbf{k}$  connects the momentum to the wavevector, which was used in Sec.~\ref{SUBSEC:TwoBody}. The momenta in Eq.~\eqref{CollisionIntegral} are related by energy and momentum conservation such that $\mathbf{p}+\mathbf{p}_1=\mathbf{p}'+\mathbf{p}_1'$ and $|\mathbf{p}-\mathbf{p}_1|=|\mathbf{p}'-\mathbf{p}_1'|$. The integration variable is $\hat{\Omega}=(\mathbf{p}'-\mathbf{p}_1')/|\mathbf{p}'-\mathbf{p}_1'|$. Equation~\eqref{EQ:KineticEq} does not provide an accurate description of the atoms within the high-phase-space-density regions, i.e., the condensates. These would presumably be described within the context of a Gross-Pitaevskii-type theory of BEC evolution~\cite{Norrie_TWA_Collisions1,Norrie_TWA_Collisions2,Deuar_Hybrid}. However, the goal of simulating both the high and low-phase-space-density components of the gas is beyond the scope of our current work. We simply wish to approximately capture the density evolution of the condensates. This in turn yields a reasonable prediction for the scattering rates and therefore the formation and deformation of the scattering halo.

We solve Eq.~\eqref{EQ:KineticEq} numerically using a DSMC algorithm, details of which have been published in Ref.~\cite{Sykes14_preprint}. We use an initial condition corresponding to both spatial and momentum densities proportional to 
\begin{equation}\label{ProbDistFun}
 P(\mathbf{R};\mathbf{x})={\rm max}\left\{
 \frac{15}{8\pi\prod_\alpha R_\alpha}\left[1-\sum_{\alpha}\frac{x_\alpha^2}{R_\alpha^2}\right],0
 \right\},\qquad \alpha = \{1,2,3\},
\end{equation}
where $\mathbf{x}$ ($R_\alpha$) can denote either spatial or momentum coordinates (widths). 
The spatial part corresponds to a Thomas-Fermi condensate density, with the widths calculated in the manner prescribed by Refs.~\cite{Eberlein2005,Griesmaier2006}. The momentum widths are found by fitting to the experimental image of the expanded condensate. The system is then divided into two halves, which propagate along the positive and negative $x$-axis, respectively. 

For a detailed account of the simulation method, we refer the reader to Ref.~\cite{Sykes14_preprint}, which adjusts the original DSMC approach~\cite{BirdBook} into a version appropriate to dipolar gases. Similar methods have been employed in the study of ultracold gases, see for instance Refs.~\cite{WadeBlakie2011_PRA,Lobo_Boltzmann,Urban_Boltzmann}. In particular, Ref.~\cite{WadeBlakie2011_PRA} also studied halo formation in the case of $s$- and $d$-wave collisional cross sections. Briefly, our computational algorithm uses test-particles with phase-space coordinates which are sampled from the initial distribution in Eq.~\eqref{ProbDistFun}. The test particles move classically from one time-step to the next. At each time-step particles are binned in position space. The bin-size represents the finite resolution of the delta-function within the numerics. Within each bin, the collision probability for each pair of particles is evaluated according to Eq.~\eqref{EQsigma} using the correct value of $\hat{k}$ corresponding to each pair. Collisions are then chosen to occur stochastically in accordance with these probabilities. The post-collision velocities are also chosen stochastically, in accordance with the differential scattering cross section in Eq.~\eqref{EQdifferential}.

\begin{figure}
\centering
 \includegraphics[width=0.7\columnwidth]{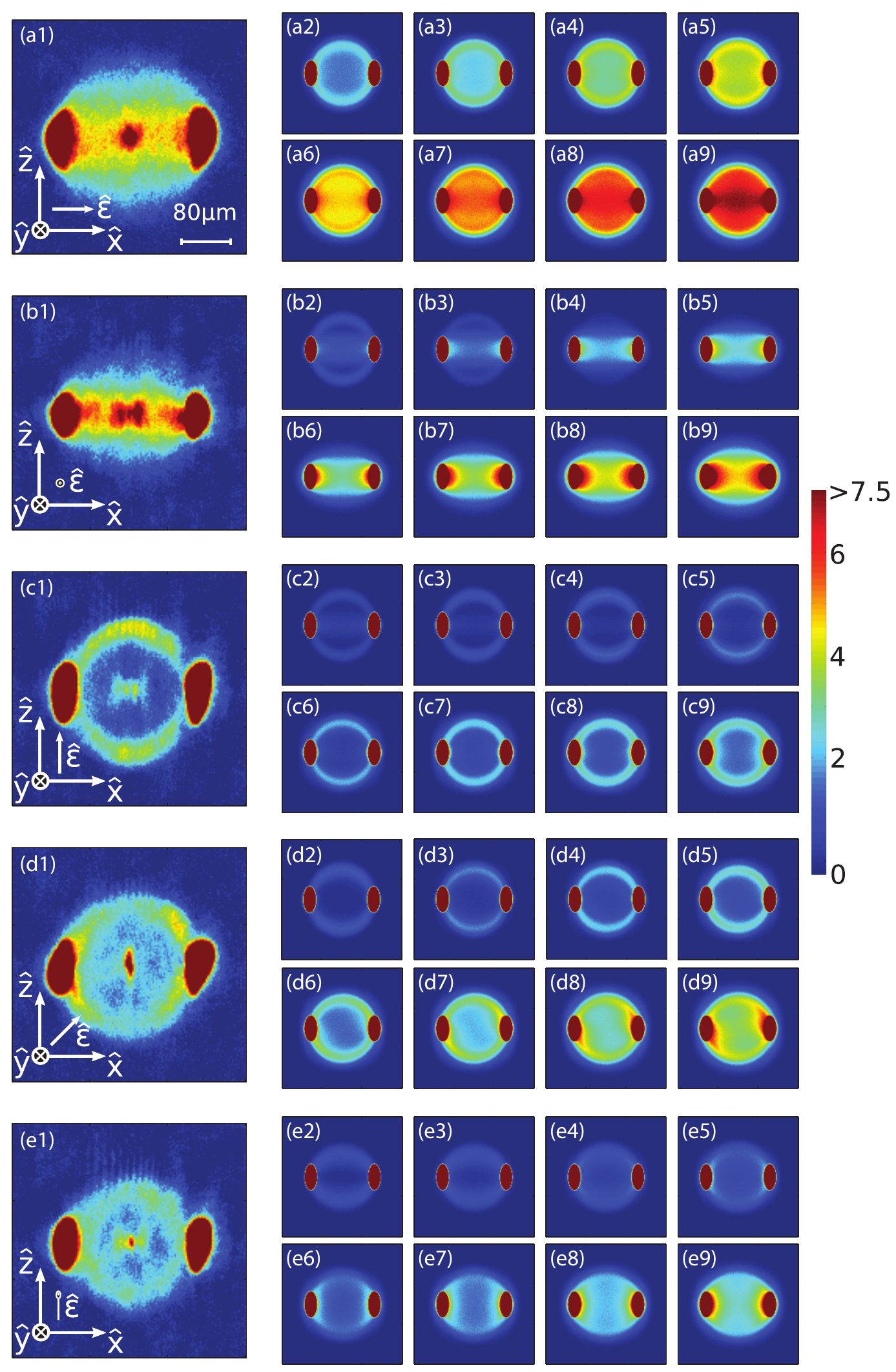}
 \caption{Absorption images in the $\hat{x}$--$\hat{z}$ plane after 22-ms TOF and averaged forty times each. The larger images labeled (a1), (b1), etc.~show the experimental data and the dipole alignment configuration, while the smaller images (a2)--(a9), (b2)--(b9), etc.~show simulation images, each with a different contribution from the $s$-wave (short-range) interaction. This contribution varies from $a=40a_0$ in (a2), (b2), etc.~to $a=180a_0$ in (a9), (b9), etc., with uniform steps of $20a_0$ in between. Dipoles are aligned along the direction: $(1,0,0)$ in (a), $(0,1,0)$ in (b), $(0,0,1)$ in (c), $(1,0,1)/\sqrt{2}$ in (d), and $(0,1,1)/\sqrt{2}$ in (e). The colour scale to the right shows the number of atoms per pixel for all data sets.}
 \label{FIG:MainComparison}
\end{figure}

\section{Results}

The experimental absorption images are presented in the first column of Fig.~\ref{FIG:MainComparison}, i.e., in panels (a1), (b1), $\ldots$ (e1). Each image corresponds to the average of forty experimental runs at a fixed dipole alignment direction, as stated in the figure caption. 
Imperfections in the diffraction leaves residual atoms near $0\hbar \mathbf{k}_{L}$. This was also observed in Ref.~\cite{QuantumNewtonsCradle}. These atoms contribute noise to the measurement both near $0\hbar \mathbf{k}_{L}$ and in the $|2\hbar \mathbf{k}_{L}|$ halos from collisions with the $\pm2\hbar \mathbf{k}_{L}$ BECs. However, the effect of the dipole alignment direction is clear from the remarkable variation between images (a1), (b1), $\ldots$ (e1).  We run a family of simulations for each value of the dipole alignment direction, but with varying $s$-wave scattering length~\footnote{We allow scattering length to vary because the large number of internal degrees of freedom within open-shell lanthanide atoms~\cite{Kotochigova_Dysprosium1,Kotochigova_Dysprosium2} restricts our ability to microscopically determine the scattering lengths of these atoms, though previous measurements have restricted the range of likely values~\cite{TangSykes2015_PRA,Maier2015,Tang2016}.}. The test particles in the DSMC are initialized as follows: We estimate the initial density of the condensate by using the exact solution to the Hartree-Fock theory for a dipolar condensate in the Thomas-Fermi limit~\cite{Eberlein2005}, and we use this to define spatial widths $R_\alpha^{\rm (spat.)}$ in Eq.~\eqref{ProbDistFun}. TOF expansion images are used to estimate the in situ momentum distribution, and we use this measurement to define momentum widths $R_\alpha^{\rm (mom.)}$ in Eq.~\eqref{ProbDistFun}. We have also looked at Gaussian distribution functions in place of Eq.~\eqref{ProbDistFun}, and also small variations in momentum and spatial widths $R_\alpha$.  However, we found that such variations have very little effect on our conclusions. These families of simulation results are shown as the smaller images in Fig.~\ref{FIG:MainComparison} panels (a2)--(a9), (b2)--(b9), $\ldots$ (e2)--(e9).

\begin{figure}[t]
\centering
 \includegraphics[width=.75\textwidth]{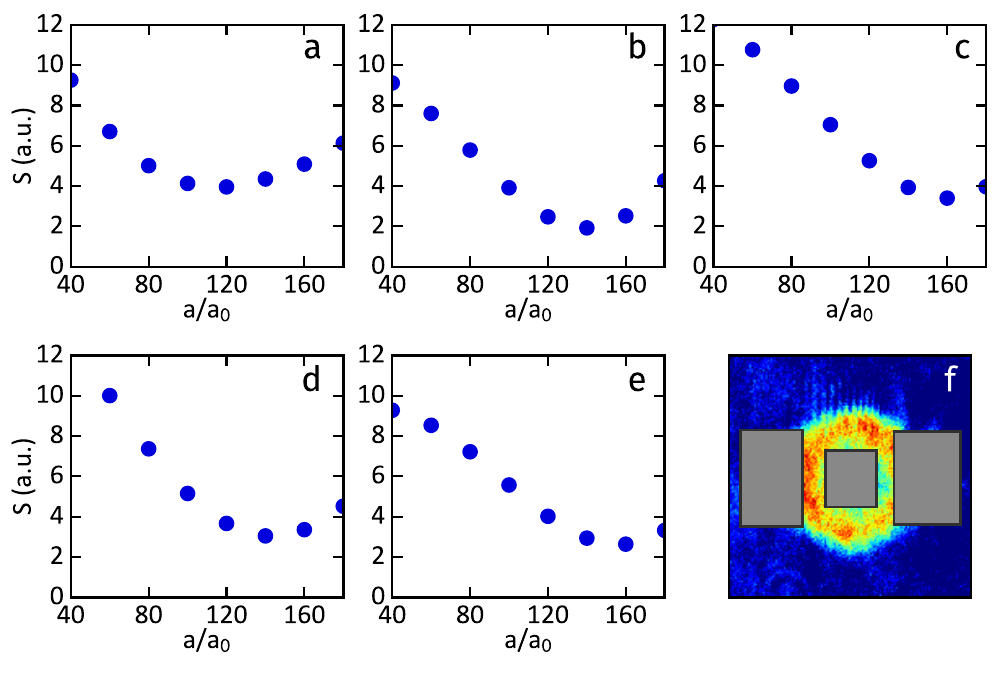}
 \caption{(a)--(e) Pixel-by-pixel comparison of the experimental and simulated absorption images for different dipole alignment directions: $(1,0,0)$ in (a), $(0,1,0)$ in (b), $(0,0,1)$ in (c), $(1,0,1)/\sqrt{2}$ in (d), and $(0,1,1)/\sqrt{2}$ in (e). The points denote the value of the weighted least squares cost function for each simulated $s$-wave scattering length $a$ (see text for details). (f) Example of the masking used for pixel-by-pixel comparisons. The grey regions to the left and right mask the unscattered BECs, and the central grey region masks atoms not diffracted by the lattice. }
 \label{FIG:PixelByPixelComparison}
\end{figure}

\begin{figure}[b]
\centering
 \includegraphics[width=.75\textwidth]{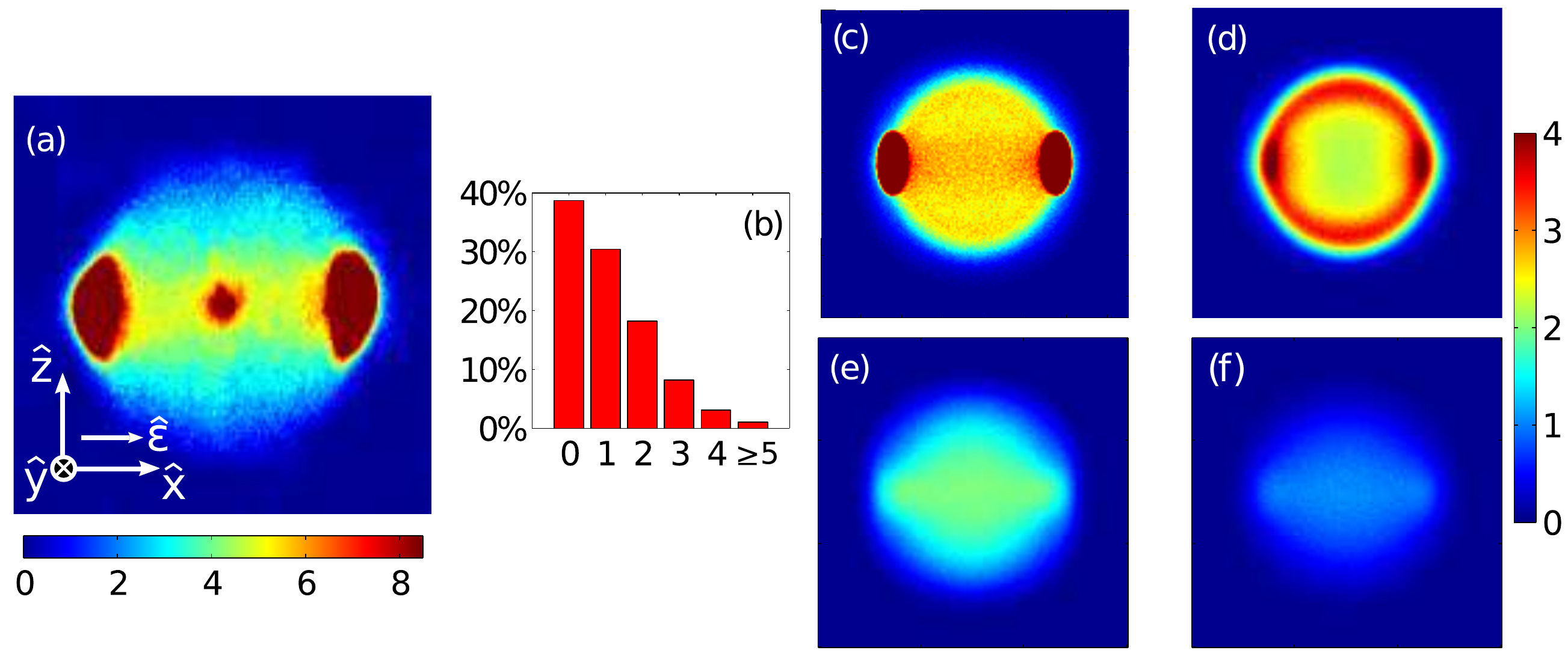}
 \caption{Analysis of multiple collision effects in the case where dipoles are aligned along the $\hat{x}$ axis.  (a) The experimental absorption image.  (b)--(f) Analysis of simulation data: (b) shows the population of atoms separated by the number of scattering events each atom incurred; (c) shows the full simulation of the absorption image; (d) an absorption image with {\it only} the atoms which scattered once, and is therefore in perfect correspondence with the two-body differential cross section; (e) shows an absorption image with only the atoms which scattered twice; (f) shows an absorption image with only the atoms which scattered three times. The simulations were done with an $s$-wave scattering length $a=140a_0$. The colour-scale (defining the number of atoms per pixel) is the same in panels (a) and (c) and shown beneath panel (a). The remaining panels correspond to the colour-scale shown on the right.
 }
 \label{FIG:MultipleCollisionsX}
\end{figure}

We perform a pixel-by-pixel comparison between the experimental and simulated scattering images. Each pixel corresponds to an area of 2.6~$\mu$m $\times$ 2.6~$\mu$m. For each dipole alignment direction, the average experimental image $\mathcal{E}$ is obtained as well as the standard error for each pixel $\mathcal{E}^{\sigma}$. The experimental image and each simulation image $\mathcal{S}$ (including unscattered atoms) is normalized to suppress the effects of atom number variations, and a mask is then applied to the images to exclude the unscattered BECs and atoms not diffracted by the lattice. An example of a masked image is shown in Fig.~\ref{FIG:PixelByPixelComparison}(f). The weighted sum of the squared residuals is calculated for each simulated image ${S=\sum_{ij}(\mathcal{E}_{ij}-\mathcal{S}_{ij})^{2}/(\mathcal{E}^{\sigma}_{ij})^{2}}$,
where the subscript $ij$ denotes the pixel in the $i$th row and $j$th column of an image. This corresponds to the cost function for a weighted least squares regression. The results of this analysis are shown in Fig.~\ref{FIG:PixelByPixelComparison}(a)--(e). Though a particular choice of scattering length, $a$, minimizes $S$ for each dipole alignment direction, the analysis is not sensitive enough to define a 1$\sigma$ confidence interval. However, as seen in Fig~\ref{FIG:PixelByPixelComparison}, all the minima fall within the range of $120a_0$--$160a_0$, which is consistent with measurements reported in our previous work~\cite{TangSykes2015_PRA,Tang2016}. We believe that the fluctuations in the location of the minima in Fig.~\ref{FIG:PixelByPixelComparison} are primarily due to an incomplete knowledge of the initial density of the BEC. Recent results suggest that dipolar interactions in strongly dipolar BECs can have surprising consequences which require beyond-mean-field effects to be quantitatively understood, see Refs.~\cite{PfauDroplet1,PfauDroplet2,SantosWachtler}. We have run additional simulations with a variety of initial densities and note that variance in initial density leads to deformations in the halo due to the additional (fewer) multiple scattering events in the case of an increased (decreased) density. These deformations are capable of shifting the minima in Fig.~\ref{FIG:PixelByPixelComparison}.

\begin{figure}[t]
\centering
 \includegraphics[width=.77\textwidth]{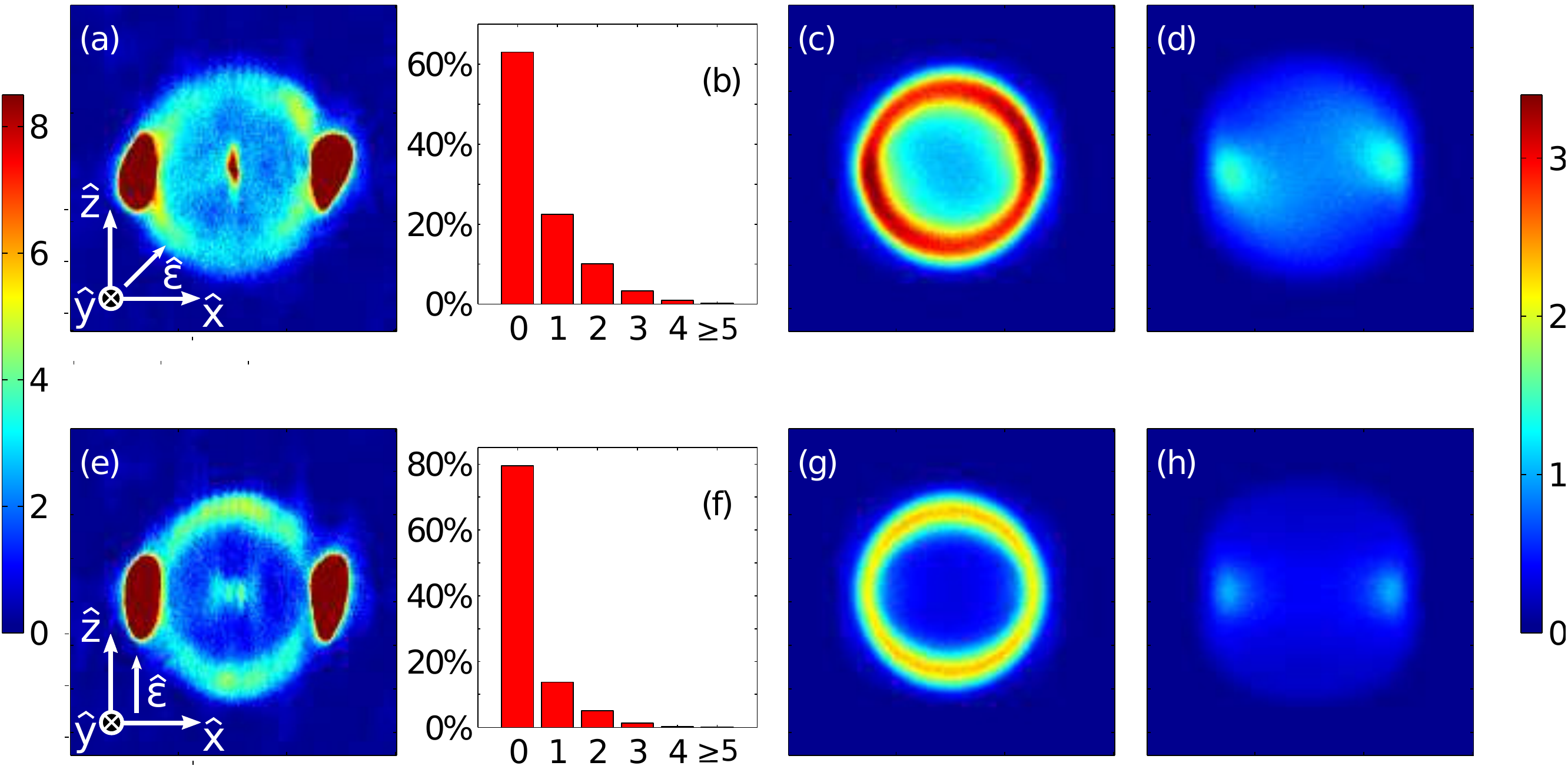}
 \caption{Similar to Fig.~\ref{FIG:MultipleCollisionsX}, except  with dipoles aligned along the $(\hat{x}+\hat{z})/\sqrt{2}$ axis [shown in the top row, (a)--(d)] and the $\hat{z}$ axis [shown in the bottom row, (e)--(h)].  (a) and (e) The experimental absorption images associated with their own colour-scale at left, indicating number of atoms per pixel.  (b)--(d) and (f)--(h) Analysis of  simulation data: (b) and (f) show the population of atoms separated by the number of scattering events each atom incurred; (c) and (g) show an absorption image with {\it only} the atoms which scattered once, and is therefore in perfect correspondence with the two-body differential cross section; (d) and (h) show an absorption image with only the atoms which scattered twice. The simulations were done with an $s$-wave scattering length $a=140a_0$. The presence of multiple collisions is less pronounced for these dipole alignment directions than in Fig.~\ref{FIG:MultipleCollisionsX}.
 }
 \label{FIG:MultipleCollisionsXZ_and_Z}
\end{figure}

We also analyze the collision frequency and, through a detailed comparison between simulation and experiment, establish the relevance of multiple collisions.  These are collisions involving atoms that have already been scattered out of the original condensates. The nature of the DSMC method allows one to label particles and keep track of collisions as they occur. 
The results are shown in Fig.~\ref{FIG:MultipleCollisionsX} for dipoles aligned along $\hat{x}$, a case in which the presence of multiple collisions is strong. Figure~\ref{FIG:MultipleCollisionsXZ_and_Z} shows results where the dipoles are aligned along both $(\hat{x}+\hat{z})/\sqrt{2}$ and $\hat{z}$, and the presence of multiple collisions is less relevant. The data can be reasonably well understood starting from Eq.~\eqref{EQsigma} and noting that, when $a=140a_0$, the total cross section has a maximum at $\eta=0^\circ$ (corresponding to dipoles aligned along $\hat{x}$) and a minimum at $\eta=90^\circ$ (corresponding to dipoles aligned anywhere in the $\hat{y}$--$\hat{z}$ plane). Specifically, we have $\sigma(0^\circ)/\sigma(90^\circ)\approx2.6$ and $\sigma(0^\circ)/\sigma(45^\circ)\approx1.6$. For this reason, we observe the unscattered fraction of the gas increases from $\sim$40\% in Fig.~\ref{FIG:MultipleCollisionsX} (b), to $\sim$60\% in Fig.~\ref{FIG:MultipleCollisionsXZ_and_Z} (b), and to $\sim$80\% in Fig.~\ref{FIG:MultipleCollisionsXZ_and_Z} (f). From this, we conclude that the experimental absorption image with dipoles aligned along the $\hat{x}$ axis is strongly affected by atoms which have undergone multiple collisions. Indeed, these atoms appear to constitute the majority of the absorption image in this case.  Reducing the initial density to suppress multiple collisions is not practicable due to the sharp loss in signal-to-noise.  However, for all other alignment directions the experimental absorption image (away from the condensate regions) is dominated by a single scattering event, and is therefore  in close correspondence with the differential scattering cross section of Eq.~\eqref{EQdifferential}.

\section{Conclusion and discussion}

In conclusion, we have measured the differential scattering between identical dipolar bosons in the low-energy (universal) regime by imaging the halos of atoms scattering from two colliding Dy BECs. The results depend strongly on the angle between the dipole alignment direction and the collision axis and are well described by the analytic formula in Eq.~\eqref{EQdifferential}, derived under the first-order Born approximation. Although it is not intended to be entirely quantitative, a classical Monte-Carlo simulation of the many-body dynamics provides a reasonable higher-order approximation to account for finite-density effects and finite-momentum distribution widths. We allow the $s$-wave scattering length to vary within the simulations, and note that discrepancies between simulation and experiment are minimized at a scattering length which remains consistent with previous measurements~\cite{TangSykes2015_PRA,Tang2016}. 

The measurements provide a beautiful demonstration of the theoretical prediction for a low energy scattering amplitude, which dates back to O'Malley in 1964~\cite{DipolarCollisions_OMalley}. The envisaged physical system under consideration at that time was not dipolar collisions, but rather the scattering of an electron by a non-spherical atom, such as atomic oxygen. However, we see that the interaction potential is asymptotically equivalent in the two cases. Future work may use such experiments to probe the complex collisional physics of dipolar condensates and degenerate Fermi gases near Feshbach resonances in and among the dense and ultradense  spectra observed in the dysprosium and erbium systems~\cite{Erbium_FermiGas,Dy_FeshbachSpectrum,Frisch2014,Maier2015Chaotic,Maier2015,Burdick:2016vv}.

\section{Acknowledgements}

We gratefully acknowledge John Bohn for early contributions to this work.  We thank Jack DiSciacca and Wil Kao for experimental assistance and Ian Spielman for a stimulating discussion.  We are grateful to the NSF and AFOSR for funding support.  Y.T.~acknowledges partial support from a Stanford Graduate Fellowship.  The research leading to these results received funding from the European  Union’s  Horizon  2020  research  and  innovation programme under grant agreement No 658311.

\medskip

\providecommand{\newblock}{}


\begin{thebibliography}{10}
\expandafter\ifx\csname url\endcsname\relax
  \def\url#1{{\tt #1}}\fi
\expandafter\ifx\csname urlprefix\endcsname\relax\def\urlprefix{URL }\fi
\providecommand{\eprint}[2][]{\url{#2}}

\bibitem{LandauQuantum}
Landau L~D and Lifshitz E~M 1958 {\em Quantum Mechanics (Non-relativistic
  Theory)\/} (Pergamon Press)

\bibitem{Fermi34_NC}
Fermi E 1934 {\em Il Nuovo Cimento\/} {\bf 11} 157

\bibitem{Bethe}
Bethe H~A and Morrison P 2006 {\em Elementary nuclear theory\/} (Courier
  Corporation)

\bibitem{Chin10_RMP}
Chin C, Grimm R, ienne P and Tiesinga E 2010 {\em Rev. Mod. Phys.\/} {\bf 82}
  1225

\bibitem{TaylorBook}
Taylor J~R 2012 {\em Scattering theory: the quantum theory of nonrelativistic
  collisions\/} (Courier Corporation)

\bibitem{PhysRevLett.80.3419}
Timmermans E and C\^ot\'e R 1998 {\em Phys. Rev. Lett.\/} {\bf 80} 3419

\bibitem{PhysRevA.59.1500}
Geist W, You L and Kennedy T~A~B 1999 {\em Phys. Rev. A\/} {\bf 59} 1500

\bibitem{DeMarcoJin1999_PRL}
DeMarco B, Bohn J~L, Burke J~P, Holland M and Jin D~S 1999 {\em Phys. Rev.
  Lett.\/} {\bf 82} 4208

\bibitem{DeMarcoJin_Science1999}
DeMarco B and Jin D~S 1999 {\em Science\/} {\bf 285} 1703

\bibitem{Pethick2002}
Pethick C~J and Smith H 2002 {\em Bose-Einstein condensation in dilute gases\/}
  (Cambridge University Press)

\bibitem{Shakeshaft_r3potential}
Shakeshaft R 1972 {\em J. Phys. B: At. Mol. Phys.\/} {\bf 5} L115

\bibitem{DipolarCollisions_OMalley}
O'Malley T~F 1964 {\em Phys. Rev.\/} {\bf 134} A1188

\bibitem{Chromium_BEC}
Griesmaier A, Werner J, Hensler S, Stuhler J and Pfau T 2005 {\em Phys. Rev.
  Lett.\/} {\bf 94} 160401

\bibitem{Dysprosium_BEC}
Lu M, Burdick N~Q, Youn S~H and Lev B~L 2011 {\em Phys. Rev. Lett.\/} {\bf 107}
  190401

\bibitem{Erbium_BEC}
Aikawa K, Frisch A, Mark M, Baier S, Rietzler A, Grimm R and Ferlaino F 2012
  {\em Phys. Rev. Lett.\/} {\bf 108} 210401

\bibitem{Dysprosium_DegenerateFermiGas}
Lu M, Burdick N~Q and Lev B~L 2012 {\em Phys. Rev. Lett.\/} {\bf 108} 215301

\bibitem{Erbium_FermiGas}
Aikawa K, Frisch A, Mark M, Baier S, Grimm R and Ferlaino F 2014 {\em Phys.
  Rev. Lett.\/} {\bf 112} 010404

\bibitem{Chromium_FermiGas}
Naylor B, Reigue A, Mar\'echal E, Gorceix O, Laburthe-Tolra B and Vernac L 2015
  {\em Phys. Rev. A\/} {\bf 91} 011603

\bibitem{KRb}
Ni K~K, Ospelkaus S, de~Miranda M~H~G, Pe{\textquoteright}er A, Neyenhuis B,
  Zirbel J~J, Kotochigova S, Julienne P~S, Jin D~S and Ye J 2008 {\em
  Science\/} {\bf 322} 231

\bibitem{RbCs}
Takekoshi T, Reichs\"ollner L, Schindewolf A, Hutson J~M, Le~Sueur C~R, Dulieu
  O, Ferlaino F, Grimm R and N\"agerl H~C 2014 {\em Phys. Rev. Lett.\/} {\bf
  113} 205301

\bibitem{NaK}
Park J~W, Will S~A and Zwierlein M~W 2015 {\em Phys. Rev. Lett.\/} {\bf 114}
  205302

\bibitem{NaRb}
Guo M, Zhu B, Lu B, Ye X, Wang F, Vexiau R, Bouloufa-Maafa N, Qu\'em\'ener G,
  Dulieu O and Wang D 2016 {\em Phys. Rev. Lett.\/} {\bf 116} 205303

\bibitem{Bohn09_NJP}
Bohn J, Cavagnero M and Ticknor C 2009 {\em New J. Phys.\/} {\bf 11} 055039

\bibitem{Bohn14_PRA}
Bohn J~L and Jin D~S 2014 {\em Phys. Rev. A\/} {\bf 89} 022702

\bibitem{Aikawa14_preprint}
Aikawa K, Frisch A, Mark M, Baier S, Grimm R, Bohn J, Jin D, Bruun G and
  Ferlaino F 2014 {\em Phys. Rev. Lett.\/} {\bf 113} 263201

\bibitem{TangSykes2015_PRA}
Tang Y, Sykes A, Burdick N~Q, Bohn J~L and Lev B~L 2015 {\em Phys. Rev. A\/}
  {\bf 92} 022703

\bibitem{swaveHalo_1995}
Gibble K, Chang S and Legere R 1995 {\em Phys. Rev. Lett.\/} {\bf 75} 2666

\bibitem{ImpurityScattering_2000}
Chikkatur A~P, G\"orlitz A, Stamper-Kurn D~M, Inouye S, Gupta S and Ketterle W
  2000 {\em Phys. Rev. Lett.\/} {\bf 85} 483

\bibitem{sanddwaveHalo_2004_Niels}
Thomas N~R, Kj\ae{}rgaard N, Julienne P~S and Wilson A~C 2004 {\em Phys. Rev.
  Lett.\/} {\bf 93} 173201

\bibitem{sanddwaveHalo_2004_Jook}
Buggle C, L\'eonard J, von Klitzing W and Walraven J~T~M 2004 {\em Phys. Rev.
  Lett.\/} {\bf 93} 173202

\bibitem{Genkina:2016cb}
Genkina D, Aycock L~M, Stuhl B~K, Lu H~I, Williams R~A and Spielman I~B 2016
  {\em New J. Phys.\/} {\bf 18} 013001

\bibitem{Niels_Nature}
Thomas R, Roberts K~O, Tiesinga E, Wade A~C~J, Blakie P~B, Deb A~B and
  Kj\ae{}rgaard N 2016 {\em Nat. Commun.\/} {\bf 7} 12069

\bibitem{PairCorrelations_MetastableHelium}
Perrin A, Chang H, Krachmalnicoff V, Schellekens M, Boiron D, Aspect A and
  Westbrook C~I 2007 {\em Phys. Rev. Lett.\/} {\bf 99} 150405

\bibitem{PairCorrelations_Theory1}
Perrin A, Savage C~M, Boiron D, Krachmalnicoff V, Westbrook C~I and Kheruntsyan
  K~V 2008 {\em New J. Phys.\/} {\bf 10} 045021

\bibitem{PairCorrelations_Theory2}
Chwede\ifmmode~\acute{n}\else \'{n}\fi{}czuk J, Zi\ifmmode~\acute{n}\else
  \'{n}\fi{} P, Trippenbach M, Perrin A, Leung V, Boiron D and Westbrook C~I
  2008 {\em Phys. Rev. A\/} {\bf 78} 053605

\bibitem{PairCorrelations_Theory3}
\"Ogren M and Kheruntsyan K~V 2009 {\em Phys. Rev. A\/} {\bf 79} 021606

\bibitem{Spielman_SyntheticPartialWaves}
Williams R~A, LeBlanc L~J, Jim{\'e}nez-Garc{\'\i}a K, Beeler M~C, Perry A~R,
  Phillips W~D and Spielman I~B 2012 {\em Science\/} {\bf 335} 314

\bibitem{Tang2015}
Tang Y, Burdick N~Q, Baumann K and Lev B~L 2015 {\em New J. Phys.\/} {\bf 17}
  045006

\bibitem{Wu2005}
Wu S, Wang Y~J, Diot Q and Prentiss M 2005 {\em Phys. Rev. A\/} {\bf 71} 043602

\bibitem{Dy_FeshbachSpectrum}
Baumann K, Burdick N~Q, Lu M and Lev B~L 2014 {\em Phys. Rev. A\/} {\bf 89}
  020701

\bibitem{DipolarBosonsReview}
Lahaye T, Menotti C, Santos L, Lewenstein M and Pfau T 2009 {\em Rep. Prog.
  Phys.\/} {\bf 72} 126401

\bibitem{Norrie_TWA_Collisions1}
Norrie A~A, Ballagh R~J and Gardiner C~W 2006 {\em Phys. Rev. A\/} {\bf 73}
  043617

\bibitem{Norrie_TWA_Collisions2}
Norrie A~A, Ballagh R~J and Gardiner C~W 2005 {\em Phys. Rev. Lett.\/} {\bf 94}
  040401

\bibitem{Deuar_Hybrid}
Deuar P 2009 {\em Phys. Rev. Lett.\/} {\bf 103} 130402

\bibitem{Sykes14_preprint}
Sykes A~G and Bohn J~L 2015 {\em Phys. Rev. A\/} {\bf 91} 013625

\bibitem{Eberlein2005}
Eberlein C, Giovanazzi S and O'Dell D~H 2005 {\em Phys. Rev. A\/} {\bf 71}
  033618

\bibitem{Griesmaier2006}
Griesmaier A 2006 {\em Dipole-dipole interaction in a degenerate quantum gas\/}
  Ph.D. thesis Stuttgart University

\bibitem{BirdBook}
Bird G~A 1994 {\em Molecular Gas Dynamics and the Direct Simulation of Gas
  Flows\/} (Clarendon)

\bibitem{WadeBlakie2011_PRA}
Wade A~C~J, Baillie D and Blakie P 2011 {\em Phys. Rev. A\/} {\bf 84} 023612

\bibitem{Lobo_Boltzmann}
Goulko O, Chevy F and Lobo C 2012 {\em New J. Phys.\/} {\bf 14} 073036

\bibitem{Urban_Boltzmann}
Pantel P~A, Davesne D and Urban M 2015 {\em Phys. Rev. A\/} {\bf 91} 013627

\bibitem{QuantumNewtonsCradle}
Kinoshita T, Wenger T and Weiss D~S 2006 {\em Nature\/} {\bf 440} 900

\bibitem{Kotochigova_Dysprosium1}
Kotochigova S and Petrov A 2011 {\em Phys. Chem. Chem. Phys.\/} {\bf 13} 19165

\bibitem{Kotochigova_Dysprosium2}
Petrov A, Tiesinga E and Kotochigova S 2012 {\em Phys. Rev. Lett.\/} {\bf 109}
  103002

\bibitem{Maier2015}
Maier T, Ferrier-Barbut I, Kadau H, Schmitt M, Wenzel M, Wink C, Pfau T,
  Jachymski K and Julienne P~S 2015 {\em Phys. Rev. A\/} {\bf 92} 060702(R)

\bibitem{Tang2016}
Tang Y, Sykes A~G, Burdick N~Q, DiSciacca J~M, Petrov D~S and Lev B~L 2016 {\em
  arXiv:1606.08856\/}

\bibitem{PfauDroplet1}
Kadau H, Schmitt M, Wenzel M, Wink C, Maier T, Ferrier-Barbut I and Pfau T 2016
  {\em Nature\/} {\bf 530} 194

\bibitem{PfauDroplet2}
Ferrier-Barbut I, Kadau H, Schmitt M, Wenzel M and Pfau T 2016 {\em Phys. Rev.
  Lett.\/} {\bf 116}(21) 215301

\bibitem{SantosWachtler}
W\"achtler F and Santos L 2016 {\em Phys. Rev. A\/} {\bf 93} 061603

\bibitem{Frisch2014}
Frisch A, Mark M, Aikawa K, Ferlaino F, Bohn J~L, Makrides C, Petrov A and
  Kotochigova S 2014 {\em Nature\/} {\bf 507} 475

\bibitem{Maier2015Chaotic}
Maier T, Kadau H, Schmitt M, Wenzel M, Ferrier-Barbut I, Pfau T, Frisch A,
  Baier S, Aikawa K, Chomaz L, Mark M~J, Ferlaino F, Makrides C, Tiesinga E,
  Petrov A and Kotochigova S 2015 {\em Phys. Rev. X\/} {\bf 5} 041029

\bibitem{Burdick:2016vv}
Burdick N~Q, Tang Y and Lev B~L 2016 {\em arXiv:1605.03211\/}

\end{thebibliography}
\end{document}